\documentclass[aps,twocolumn,showpacs]{revtex4}
\usepackage{graphicx}   %       for graphics
\usepackage{latexsym}   %       for special symbols

\newcommand{\beq}{\begin{equation}}
\newcommand{\eeq}{\end  {equation}}

\newcommand{\beqar}{\begin{eqnarray}}
\newcommand{\eeqar}{\end  {eqnarray}}

\newcommand{\bold}[1]{\mbox{\boldmath $#1$}}    %       bold symbol
\newcommand{\smbold}[1]{\mbox{\boldmath\scriptsize $#1$}}%   small bold char
\newcommand{\tnbold}[1]{\mbox{\boldmath\tiny $#1$}}%   tiny bold char
\newcommand{\zero}[1]{\hspace{0.1ex}\raisebox{1.0ex}{\scriptsize {$\circ$}}
  \hspace{-1ex}{#1}}
\newcommand{\ie}{{\em i.e.}}			%	i.e.
\newcommand{\del}{\partial}                     %       partial
\newcommand{\GeV}{{\rm GeV}}			%	GeV
\newcommand{\MeV}{{\rm MeV}}                    %       MeV
\newcommand{\fm}{{\rm fm}}                      %       fm
\newcommand{\eps}{\varepsilon}
\newcommand{\bfk}{\bold{k}}			%       bold math k
\newcommand{\smk}{{\smbold{k}}} 		%       small bold math k
\newcommand{\tnk}{{\tnbold{k}}}			%	tiny bold math k
\newcommand{\bfr}{\bold{r}}			%       bold math r
\newcommand{\smr}{{\smbold{r}}} 		%       small bold math r
\newcommand{\bfv}{\bold{v}}			%       bold math v
\newcommand{\rme}{{\rm e}}                      %       roman e
\newcommand{\grad}{\bold{\nabla}}
\newcommand{\ihalf}{\mbox{${i\over2}$}}		%	i/2
\newcommand{\half}{\mbox{${1\over2}$}}          %       1/2
\newcommand{\third}{\mbox{${1\over3}$}}         %       1/3
\newcommand{\twothird}{\mbox{${2\over3}$}}	%	2/3
\newcommand{\fourthird}{\makebox{$4\over3$}}	%	4/3
\newcommand{\threehalf}{\makebox{$3\over2$}}	%	3/2
\begin{document}
%==============================================================================
~\

\title{Spinodal phase separation in relativistic nuclear collisions}

\author{J{\o}rgen Randrup}

\affiliation{Nuclear Science Division, Lawrence Berkeley National Laboratory,
Berkeley, California 94720, USA}

\date{July 1, 2010}	%\date{\today}

\begin{abstract}
The spinodal amplification of density fluctuations is treated perturbatively
within dissipative fluid dynamics for the purpose of elucidating
the prospects for this mechanism to cause a phase separation
to occur during a relativistic nuclear collision.
The present study includes not only viscosity but also heat conduction
(whose effect on the growth rates is of comparable magnitude but opposite),
as well as a gradient term in the local pressure,
and the corresponding dispersion relation for collective modes 
in bulk matter is derived from relativistic fluid dynamics.
A suitable two-phase equation of state is obtained 
by interpolation between a hadronic gas and a quark-gluon plasma,
while the transport coefficients are approximated by simple parametrizations
that are suitable at any degree of net baryon density.
We calculate the degree of spinodal amplification
occurring along specific dynamical phase trajectories
characteristic of nuclear collision at various energies.
The results bring out the important fact that the prospects for
spinodal phase separation to occur can be greatly enhanced
by careful tuning of the collision energy to ensure that the
thermodynamic conditions associated with the maximum compression
lie inside the region of spinodal instability.
\end{abstract}


\maketitle

%==============================================================================
\section{Introduction}

It is expected that the confined and deconfined phases of
strongly interacting matter may coexist 
at net baryon densities above a certain critical value
and significant experimental efforts are underway to search for
evidence of the associated first-order phase transition
and its critical end point:
a systematic beam-energy scan is currently being performed at RHIC (BNL)
to look for the critical point \cite{RHIC-BES};
the CBM experiment at FAIR (GSI) will study baryon-dense matter
and search for the phase transition \cite{CBM-book};
and the proposed NICA (JINR) aims at exploring the mixed phase \cite{NICA}.

These studies are rather challenging,
not only because it is inherently difficult to extract the properties
of equilibrium bulk matter from collision experiments,
but also because there is currently no suitable transport model available
for guiding these efforts.
As a result, there is currently an urgent need for identifying
experimentally observable signals of the phase structure.

The present paper focusses on the possibility that 
the mechanism of spinodal phase decomposition
may have effects that could be exploited as signals of the phase transition.
Spinodal decomposition is a well-known generic phenomenon
associated with first-order phase transitions 
that has been studied in a variety of substances
and also found industrial application \cite{spinodal}.
Furthermore, nuclear spinodal fragmentation \cite{PhysRep389} was observed 
in nuclear collisions at intermediate energies \cite{BorderiePRL86}
several years ago.
A preliminary study \cite{RandrupPRC79} found grounds for 
guarded optimism that spinodal separation between the confined and deconfined
phases could in fact occur in relativistic collisions
and we have therefore undertaken the present more refined analysis.

While that earlier study \cite{RandrupPRC79} employed a somewhat schematic 
equation of state based on a generalized classical gas,
the present uses a more realistic equation of state obtained
by interpolating between a hadron gas and a quark-gluon plasma.
An advantage of this procedure is that it automatically incorporates the
increase in the number of degrees of freedom in the dense (deconfined) phase, 
a peculiar but important characteristic of strongly interacting matter.
Building on the developments in Ref.\ \cite{RandrupPRC79},
we take account of finite-range effects
by including a gradient term in the equation of state.
This refinement is essential 
for obtaining a physically meaningful description
since it ensures both that there is an interface tension between
the two coexisting phases 
and that the spinodal growth rates subside at large wave numbers.

We again carry out our studies within the framework of fluid dynamics,
because this type of transport description has the distinct advantage that 
the complicated and still poorly understood microstructure of the system
enters only via the equation of state and the transport coefficients.
A general discussion of the  fluid-dynamical description of \
first-order phase transitions was given recently in Ref.\ \cite{SkokovNPA828}.

Although the dispersion equation in \cite{RandrupPRC79}
was derived with both shear and bulk viscosity included, 
the actual calculations were done for ideal fluid dynamics.
In the present study, the dynamical calculations include
not only viscosity but also heat conductivity,
which proves to be as important as viscosity while affecting 
the spinodal growth rates oppositely, as we demonstrate quantitatively.
The associated cubic dispersion equation 
is derived directly from relativistic fluid dynamics.
The medium dependence of the transport coefficients 
is expressed in terms of the local values of 
the enthalpy density and the particle spacing,
an approximation that applies not only in the baryon-poor regime
but also in the baryon-dense media of relevance to the phase transition.
The strength of the transport coefficients for arbitrary
density and temperature can thus be related to the values
obtained from analyses of the RHIC data.

We seek to construct plausible dynamical phase trajectories
by invoking results from earlier calculations with various transport models
and we examine in particular the crucial importance of 
using a collision energy for which the maximum compression
occurs inside the spinodal phase region.
Once the phase trajectory of the collision system has been specified,
we may integrate the spinodal growth rate along the dynamical history
and thus calculate the resulting degree of amplification
as a function of the wave number of the density perturbation.
We do that for a range of dissipation strengths that bracket those
expected from the analysis of the RHIC data.

The present, more refined, studies suggest that spinodal phase decomposition
may indeed be triggered during nuclear collisions 
within a certain (likely relatively narrow) optimal energy range.
This expectation is rather insensitive to the still poorly known strength 
of the transport coefficients.
Such a spinodal phase separation would result in an assembly of
plasma drops embedded in a hadron gas 
and our present analysis permits us to estimate the typical drop size.

The presentation is organized as follows.
We first discuss the expected thermodynamic phase structure
of strongly interacting matter within the framework of the specific 
equation of state that we have constructed.
We then turn to dissipative fluid dynamics within which we derive 
the dispersion equation for the collective modes in bulk matter.
Subsequently we develop expressions for the transport coefficients 
in baryon-rich matter and use those in calculations of the 
spinodal growth rates.
Finally, we obtain quantitative results 
for the degree of spinodal amplification
experienced by the bulk of the collision system 
as it evolves along various plausible phase trajectories.
The construction of the equation of state
and the associated spline procedure are described in appendices.

%==============================================================================
\section{Thermodynamic phase structure}
\label{thermo}

In order to make quantitative studies, we need to employ a specific
equation of state that is plausibly realistic and, in particular,
exhibits the expected phase structure.

\begin{figure}          %       -----------------------------------------
\includegraphics[angle=0,width=3.1in]{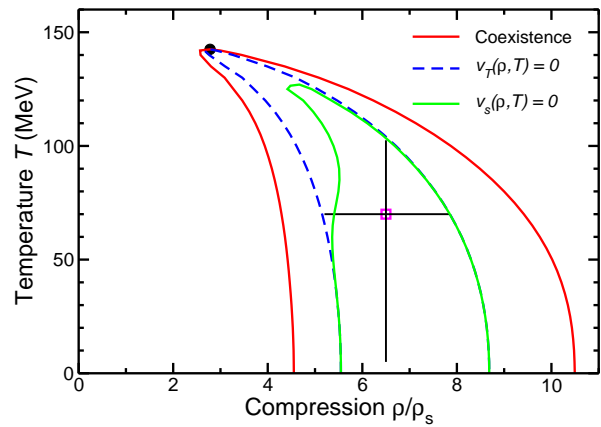}	%	phase crossing
\caption{The $(\rho,T)$ phase diagram for bulk matter
showing the phase coexistence boundary (outer solid),
the isothermal spinodal where $v_T=0$ (dashed), 
and the isentropic spinodal where $v_s=0$ (inner solid),
together with the critical point (for details, see App.\ \ref{EoS}).
The dispersion relation shown in Fig.\ \ref{f:gamma} 
was calculated at the square,
while the results shown in Fig.\ \ref{f:gmax}
were obtained along the two straight lines.
}\label{f:T-c}
\end{figure}

Although significant progress has been made in understanding the 
thermodynamic properties of the confined and deconfined phases separately, 
our current understanding of the phase coexistence region 
is not yet on firm ground.
We therefore employ a conceptually simple approximate equation of state 
in which the region of phase transformation is described
by means of a suitable interpolation between an idealized hadron gas
and an idealized quark-gluon plasma.
The details of this construction are described in Appendix \ref{EoS},
while the resulting phase structure is shown in Fig.\ \ref{f:T-c}.

            %       -----------------------------------------

For the present study, it is convenient to work in the canonical representation
where the thermodynamic state of the system is characterized by the
temperature $T$ and the (net) baryon density $\rho=n_B-n_{\bar B}$.
(In the phase region of primary interest, the temperature is relatively low
and the chemical potential relatively high, 
so the equilibrium population of antibaryons is relatively insignificant,
$n_{\bar B}\ll n_B$.)
The key thermodynamic function is then the free energy density,s $f_T(\rho)$,
from which the other quantities may be obtained,
\beqar
&~& {\rm Chemical~potential:}\,\,\ \mu_T(\rho)\ =\ \del_\rho f_T(\rho)\ ,\\
&~& {\rm Pressure:}\,\,\ p_T(\rho)\ =\ \rho\del_\rho f_T(\rho)-f_T(\rho)\ ,\\
&~& {\rm Entropy~density:}\,\,\ \sigma_T(\rho)\ =\ -\del_T f_T(\rho)\ ,\\
&~& {\rm Energy~density:}\,\,\ \eps_T(\rho)\ 
	=\ f_T(\rho)-T\del_T f_T(\rho)\ .\,\,\,\
\eeqar
We may also express the isothermal sound speed $v_T$,
\beq\label{vT2}
v_T^2\ \equiv\ {\rho\over h_T} \left({\del p\over\del\rho}\right)_T\
=	{\rho\over h_T} \del_\rho p_T(\rho)\ 
=\	{\rho^2\over h_T}\del_\rho^2 f_T(\rho)\ .
\eeq
where $h_T(\rho)=p_T(\rho)+\eps_T(\rho)$ is the enthalpy density,
as well as the isentropic sound speed $v_s$,
\beq
v_s^2\ \equiv\ {\rho\over h_T} \left({\del p\over\del\rho}\right)_s\
=\	v_T^2+{T\over h_T}{1\over c_v} (\del_T p_T(\rho))^2\ \geq\ v_T^2\ ,
\eeq
where the specific heat is
$c_v\equiv\del_T\eps_T(\rho)=T\del_T\sigma_T(\rho)$.

Different manifestations of the system that have the same values of
temperature, chemical potential, and pressure are in mutual thermodynamic
equilibrium and may thus coexist.
As Fig.\ \ref{f:T-c} shows, such phase coexistence 
occurs for temperatures below the critical value, $T<T_c\approx142.5$~MeV.
The corresponding density values, $\rho_1^T$ and $\rho_2^T$, are traced out
(outer solid curve).  
Inside this boundary, it is thermodynamically favorable for uniform matter
to separate into the two coexisting phases.
However, in the part of the phase coexistence region that is close to 
the boundary any small deviation from uniformity is thermodynamically 
unfavorable and here the system is mechanically metastable, 
as signalled by the fact that $v_T^2>0$ in this phase region.

This situation changes radically at the spinodal boundary
where the isothermal sound speed vanishes, $v_T=0$
(the dashed curve in Fig.\ \ref{f:T-c}):
Inside this boundary even infinitesimal deviations from uniformity are
thermodynamcially favorable and bulk matter is thus mechanically unstable.
As a consequence, small density fluctuations may be amplified and thereby
cause the system to undergo a spontaneous phase separation.
The phase region of spinodal instability extends in temperature 
all the way from zero up to $T_c$.

Finally, Fig.\ \ref{f:T-c} also shows the boundary where the isentropic
sound speed vanishes, $v_s=0$. Inside this smaller phase region,
the spinodal amplification process can occur without any entropy production.
This special region does not extend all the way up to $T_c$ and it narrows
considerably as the temperature approaches its upper bound.

The equation of state applies to idealized uniform matter.
However, since we are interested in following the evolution
of density disturbances, it is essential to include finite-range effects.
Indeed, the physical coexistence between two different phases along 
a common interface could not be realized without taking proper account of 
the density gradients, nor could the associated interface tension be obtained.
We shall therefore augment the bulk thermodynamics
with a gradient term as proposed in Ref.\ \cite{RandrupPRC79}
(see Sect.\ \ref{grad}),
which then extends the validity of the equation of state 
to non-uniform systems.
In particular, it is possible to describe the diffuse interface
between two coexisting phases and the associated tension.
Furthermore, as we shall see, the gradient term is essential
for obtaining a physically reasonable dispersion relation
because it stabilizes disturbances having a small spatial scale.

%==============================================================================
\section{Dissipative fluid dynamics}

We wish to employ dissipative fluid dynamics for our dynamical studies.
Fluid dynamics is convenient because the specific miscroscopic 
structure of the matter under consideration enters only via the equation
of state and a few transport coefficients.
On the other hand, the treatment relies on the assumption of approximate
local equilibrium which may generally be questionable in nuclear collisions.
Fortunately, our applications are to collisions of relatively 
modest energies and, moreover, to the later stages in the evolution.
Thus the conditions for applicability should be reasonably favorable.

We base our treatment on the relativistic formulation by
Muronga \cite{MurongaPRC76a}.
The four-velocity of the local flow is $u^\mu=(\gamma,\gamma\bfv)$
and the symmetric tensor $\Delta^{\mu\nu}\equiv g^{\mu\nu}-u^\mu u^\nu$
projects onto the 3-space orthogonal to $u^\mu$, $\Delta^{\mu\nu}u_\nu=0$.\\
The space-time derivative decomposes, $\del^\mu=u^\mu D+\nabla^\mu$,
where the convective time derivative is
$D\equiv u^\mu\del_\mu	%=u^0\del_0+u^i\del_i
	=\gamma[\del_t+\bfv\cdot\grad]$
while the gradient is $\nabla^\mu\ \equiv\ \Delta^{\mu\nu}\del_\nu$.

The equations of motion simplify when expressed in a specific reference frame.
For the present study, it is convenient to use the Eckart frame,
which is defined in terms of the charge flow
and is usually employed in non-relativistic scenarios.
Then, since the local charge flow, $V^\mu$, vanishes by definition,
the charge four-current density is $N^\mu=\rho u^\mu$,
where $\rho$ is the charge density in the local flow frame.
However, it may generally be preferable to use the Landau frame 
because it is defined in terms of the energy flow and is thus meaningful 
also for chargeless fluids and fluids with several conserved charges 
for which there is no unique generalization of the Eckart frame.

%------------------------------------------------------------------------------
\subsection{Small disturbances}

We consider the early evolution of small deviations from uniformity.
We generally assume that these are planar and harmonic,
$\rho(\bfr,t)=\rho_0+\rho_k\exp(ikx-i\omega t)$ 
with $\delta\equiv|\rho_k|/\rho_0\ll1$, and similarly for the other quantities.
We may then ignore terms of order ${\cal O}(\delta^2)$ and higher.
It follows that the associated flow velocities are small, $v\ll1$
since ${\cal O}(v)={\cal O}(\delta)$, and thus we have 
$\gamma\equiv[1-v^2]^{-1/2}=1+{\cal O}(\delta^2)\approx1$.

Generally the energy flow is given by $W^\mu\equiv q^\mu+hV^\mu=q^\mu$,
where $q^\mu$ is the heat flow and $h=p+\eps$ the enthalpy density.
Since the local charge flow $V^\mu$ vanishes when we use the Eckart frame
(see above), the energy flow equals the heat flow, $W^\mu=q^\mu$,
and is ${\cal O}(\delta)$.  Hence ${\cal O}(Wu)={\cal O}(\delta^2)$,
and the energy-momentum tensor simplifies to
\beq
T^{\mu\nu}\ =\ 
	\eps u^\mu u^\nu -p\Delta^{\mu\nu}+\pi^{\mu\nu}-\Pi\Delta^{\mu\nu}\ .
\eeq
Here $\eps=u_\mu T^{\mu\nu} u_\nu$ is the energy density 
in the local flow frame
and $p+\Pi=-\third\Delta_{\mu\nu} T^{\mu\nu}$ is the sum of the local isotropic
pressure $p$ and the pressure induced by the bulk viscosity
which enters through the bulk pressure,
\beq
\Pi\ =\ -\zeta\nabla_\mu u^\mu\ \approx\ -\zeta\nabla_i v^i\
=\ -\zeta\del_iv^i\ =\ %-\zeta\del v^i/\del x^i\ =\
-\zeta\grad\cdot\bfv\ .
\eeq
Furthermore, the heat flow is $q^\mu=u_\nu T^{\nu\lambda}\Delta_\lambda^\mu$,
while the shear viscosity enters via the stress tensor
\beqar
\pi^{\mu\nu} %&=& T^{<\mu\nu>}\ =\ 2\eta\nabla^{<\mu}u^{\nu>}\\
&=& \eta[\Delta^\mu_\sigma\Delta^\nu_\tau + \Delta^\mu_\tau\Delta^\nu_\sigma
-\twothird\Delta^{\mu\nu}\Delta_{\sigma\tau}]\nabla^\sigma u^\tau\\
&\approx& \eta[\Delta^\mu_i\Delta^\nu_j + \Delta^\mu_j\Delta^\nu_i
-\twothird\Delta^{\mu\nu}\Delta_{ij}]\nabla^i v^j\ ,
\eeqar
where we have used that only the spatial components of $\nabla^\sigma$
contribute to leading order in $\delta$ and, moreover,
any derivatives of $u^0\approx1$ can be ignored 
so only the spatial components $u^i\approx v^i$ contribute.
Furthermore, since $\Delta^0_i=v^i$, which is ${\cal O}(\delta)$,
only the spatial elements $\Delta^i_j\approx\delta_{ij}$ contribute.
Hence only the $3\times3$ spatial part $\bold{\pi}$ is non-vanishing.
It has the following elements,
\beq
\pi^{ij}\ \approx\ -\eta[\del_iv^j+\del_jv^i-\twothird\delta_{ij}\del_kv^k]\ .
\eeq

Thus, for small deviations from uniformity,
the  spatial part $\bold{T}$ of the energy-momentum tensor is given by
\beq
T^{ij}\ \approx\ \delta_{ij}p
-\eta[\del_iv^j+\del_jv^i-\twothird\delta_{ij}\del^kv^k]
	-\zeta\delta_{ij}\grad\cdot\bfv\ .
\eeq
If the spatial variation of the viscosity coefficients $\eta$ (shear)
and $\zeta$ (bulk) may be ignored, we have
\beq\label{gradT}
\grad\cdot\bold{T}\ \approx\
\grad p-\eta\bold{\Delta}\bfv-[\third\eta+\zeta]\grad(\grad\cdot\bfv)\ ,
\eeq
The gradient simplifies further in a semi-infinite geometry 
(where the only spatial variation is in the $x$ direction),
\beq\label{dTdx}
\grad\cdot\bold{T}\ \asymp\
\del_x T_{xx}\ \approx\ \del_xp-[\fourthird\eta+\zeta]\del_x^2v\ .
\eeq
Thus only the {\em effective} viscosity
$\xi\equiv\fourthird\eta+\zeta$ enters.
It follows that a uniform stretching, $v(x)\sim x$, is dissipation free.
We also wish to point out that the shear viscosity contributes 
even though the flow has no shear.

It is interesting to note that the above result reflects %brings out 
a general
feature of isotropic expansions in $N$ dimensions.
To see this, 
assume that $\rho(\bfr)=\rho(r)$ and $\bfv(\bfr)=v(r)\bold{\hat{r}}$.
The viscous term in the Euler equation may then be evaluated
by use of spherical coordinates,
\beq
\eta\bold{\Delta}\bfv+[\third\eta+\zeta]\grad(\grad\cdot\bfv)\ =\
\xi\,\hat{\bfr}\, \del_r{1\over r^{N-1}}\del_rr^{N-1}v\ .
\eeq
Thus it is always the combination $\xi\equiv\fourthird\eta+\zeta$ that enters.
Furthermore, it follows that a Hubble-type expansion, $v(r)\sim r$, 
is dissipation free in any dimension.

%------------------------------------------------------------------------------
\subsection{Equations of motion}

The fluid-dynamic equations of motion reflect the conservation of
(baryon) charge, momentum, and energy.
We are interested in the dynamics of small deviations 
from uniformity in a semi-infinite configuration 
and we focus on harmonic disturbances,
\beqar\label{rho}
\rho(\bfr,t) &=& \rho_0+\delta\rho(x,t)\
\doteq\ \rho_0+\rho_k\rme^{ikx-i\omega t}\ ,\\
\eps(\bfr,t) &=& \eps_0+\delta\eps(x,t)\
\doteq\ \eps_0+\eps_k\rme^{ikx-i\omega t}\ ,\\
p(\bfr,t) &=& p_0+\delta p(x,t)\
\doteq\ p_0+p_k\rme^{ikx-i\omega t}\ ,
\eeqar
and similarly for the other dynamical variables.

The conservation of charge is ensured by the continuity equation,
$\del_\mu N^\mu\doteq0$, which here becomes
\beq
C:\,\ \del_t\rho\ \doteq\ -\rho_0\del_xv\,\ \Rightarrow\,\
\omega\rho_k\ \doteq\ \rho_0kv_k\ .
\eeq
It serves to eliminate the flow velocity,
$v_k=\omega\rho_k/(\rho_0k)$.
The momentum equation simplifies considerably for the present scenario
of small disturbances,
\beq\label{M}
M:\,\ h_0\del_tv\ \doteq\ -\del_x[p-\zeta\del_xv]
-\del_x\pi_{xx}-\del_tq\ ,
\eeq
where $h_0=p_0+\eps_0$ is the enthalpy density of the uniform system
and the heat flow is  $\bold{q}=(q,0,0)$ (see below).
The equation for energy conservation is similarly simplified,
\beq\label{E}
E:\,\ \del_t\eps\ \doteq\ -h_0\del_xv-\del_xq\ .
\eeq
By combining these latter two equations, 
(\ref{M}) and (\ref{E}), one obtains the sound equation,
\beq\label{sound}
\del_t{E}-\del_xM:\,\ \del_t^2\eps\ \doteq\ 
\del_x^2{\Delta}[p-\zeta\del_xv]+\del_x^2\pi_{xx}\ ,
\eeq
which amounts to
$ \omega^2\eps_k\ \doteq\ k^2p_k-i\xi (\omega/\rho_0) k^2\rho_k$
where we recall that $\xi\equiv\fourthird\eta+\zeta$,
see Eq.\ (\ref{dTdx}).

%------------------------------------------------------------------------------
\subsection{Dispersion equation}

When heat conductivity is ignored ($\kappa=0$),
the energy density tracks the charge density,
as follows immediately from the energy equation,
$\rho_0\eps_k\doteq h_0\rho_k$.
Furthermore, in the absence of a gradient term in the equation of state
(see later), we have $p_k=p_\eps\eps_k+p_\rho\rho_k$ with 
$p_\eps\equiv\del_\eps p_0(\eps,\rho)$ and 
$p_\rho\equiv\del_\rho p_0(\eps,\rho)$
where $p_0(\eps,\rho)$ is the microcanonical equation of state.
Since the isentropic sound speed $v_s$ is given by
$v_s^2=p_\eps+(\rho_0/h_0)p_\rho$,
we obtain the familiar viscous dispersion equation,
 $\omega^2=v_s^2k^2-i\xi(\omega/h_0)k^2$ \cite{RandrupPRC79}.

To obtain the dispersion equation with heat conductivity included,
we must invoke the form of the heat current,
\beq
q \approx -\kappa[\del_xT+T_0\del_tv]:\,\
q_k %=\ -i\kappa[kT_k-T_0\omega v_k]\ 
= -i\kappa[kT_k-{T_0\over\rho_0}{\omega^2\over k}\rho_k]\ .
\eeq
Insertion of this expression
into the energy equation (\ref{E}) yields a relationship between
$\rho_k$, $\eps_k$, and $T_k$,
\beq
\eps_k\ =\ {h_0\over\rho_0}\rho_k+{k\over\omega}q_k\
%=\ {h_0\over\rho_0}\rho_k-i\kappa{k^2\over\omega}T_k 
%+i\kappa{T_0\over\rho_0}\omega\rho_k\
\approx\ {h_0\over\rho_0}\rho_k-i\kappa{k^2\over\omega}T_k\ ,
\eeq
where the term $\sim\kappa\rho_k$ has been ignored 
because if is ${\cal O}(\kappa)$
in comparison with $(h_0/\rho_0)\rho_k$.
Thermodynamics enables us to express $\delta\eps$ 
in terms of $\delta T$ and $\delta\rho$,
\beq\label{epsk}
\eps_k = \left({\del\eps\over\del T}\right)_\rho T_k
	+\left({\del\eps\over\del\rho}\right)_T \rho_k\
=\	-c_v T_k	%-{1\over T^2\sigma_{\eps\eps}}T_k
-	{\sigma_{\eps\rho}\over\sigma_{\eps\eps}}\rho_k\ ,
\eeq
where $\sigma_{\eps\eps}\equiv\del_\eps^2\sigma_0(\eps,\rho)$
and $\sigma_{\eps\rho}\equiv\del_\eps\del_\rho\sigma_0(\eps,\rho)$
are second derivatives of the entropy density $\sigma_0(\eps,\rho)$.
We have also used $(\del\eps/\del T)_\rho=c_v=-1/T\sigma_{\eps\eps}$,
the heat capacity at constant density.
Using furthermore
\beq
h_0\sigma_{\eps\eps}+\rho_0\sigma_{\rho\eps}
=\ -\left({\del p\over\del\eps}\right)_\rho\ ,
\eeq
we may then obtain $T_k$ from the energy equation (\ref{E}),
\beq\label{Tk}
T_k\ \approx\  {1\over1+i\kappa k^2/\omega c_v}\
{T_0\over\rho_0}\left({\del p\over\del\eps}\right)_\rho\rho_k\ .
\eeq

The canonical equation of state $p_T(\rho)$ allows us to express
the pressure variation in terms of the variations in temperature and density,
\beq\label{pk}
p_k = \left({\del p\over\del T}\right)_\rho\!\! T_k
+	\left({\del p\over\del\rho}\right)_T\!\! \rho_k
=	\left({\del p\over\del \eps}\right)_\rho\!\! c_v T_k
+	{h_0\over\rho_0}\,v_T^2\,\rho_k ,
\eeq
where $v_T$ is the isothermal sound speed, see Eq.\ (\ref{vT2}).
%$v_T^2=(\rho/h)(\del_\rho p)_T$.

With these preparations, the dispersion equation can then be obtained
by substituting the relations (\ref{epsk}), (\ref{Tk}), (\ref{pk}) 
into the sound equation (\ref{sound}) and using the relationship
$v_s^2-v_T^2=(T/h)(\del p/\del T)_\rho(\del p/\del\eps)_\rho$,
\beq
\omega^2\ \doteq\ v_T^2k^2 -i\xi{\omega\over h_0}k^2
+	{v_s^2-v_T^2\over1+i\kappa k^2/\omega c_v}k^2\ ,
\eeq
retaining heat conduction terms only up to ${\cal O}(\kappa)$.
This is recognized as the dispersion equation given in 
Ref.\ \cite{HeiselbergAP223}.

%----------------------------------------------------------------------------
\subsubsection{Gradient correction}
\label{grad}

As noted above (end of Sect.\ \ref{thermo}),
it is essential to take account of finite-range effects,
without which the spinodal growth rate would become ever larger
as the wave number is increased \cite{RandrupPRL92}.
Following Ref.\ \cite{RandrupPRC79},
we introduce a gradient correction in the equation of state.
To leading order in the disturbance amplitudes,
the effect of the gradient term on the local pressure is given by
\beq
p(\bfr)\ \approx\ p_0(\eps(\bfr),\rho(\bfr))
	-C\rho_0\nabla^2\rho(\bfr)\ ,
\eeq
where $p_0(\eps,\rho)$ is the microcanonical equation of state,
{\em i.e.}\ the pressure in uniform matter having the specified 
energy and charge densities.
The pressure amplitude is then modified accordingly,
\beq
p_k\ \leadsto\ p_k+C\rho_0 k^2\rho_k\ .
\eeq
Hence we should augment the $\rho_k$ term in Eq.\ (\ref{pk}),
\beq
{h_0\over\rho_0}v_T^2\rho_k\ \leadsto\ 
[{h_0\over\rho_0}v_T^2+C\rho_0 k^2]\,\rho_k\ .
\eeq
The full dispersion equation is then
\beq\label{omega2}
\omega^2\ \doteq\ v_T^2k^2 +C{\rho_0^2\over h_0} k^4
	-i\xi{\omega\over h_0}k^2
+	{v_s^2-v_T^2\over1+i\kappa k^2/\omega c_v}k^2\ ,
\eeq
\ie\ the gradient term $\sim Ck^4$ is simply added, 
just as when there is no heat conductivity \cite{RandrupPRC79}.

%----------------------------------------------------------------------------
\subsubsection{Solution of the dispersion equation}

When heat conductivity is included, $\kappa>0$,
the dispersion equation is of third order and, consequently,
there are three eigenmodes for each wave vector $\bfk$,
as one would generally expect since the energy density $\eps$
is now no longer tied to the baryon density $\rho$.
This equation has one purely imaginary solution, 
$\omega_\smk^0=i\gamma_k^0$,
and a pair of generally complex solutions, $\omega_\smk^\pm$,
which are either both also imaginary, $\omega_\smk^\pm=i\gamma_k^\pm$,
or have the form $\omega_\smk^\pm=\mp\epsilon_k+i\gamma_k$.
In the latter case it is easy to see that $\gamma_\smk<0$ 
in the normal region where $v_T^2>0$.

The two frequencies $\zero{\omega}_k^\pm$ obtained for ideal 
(\ie\ non-dissipative) fluid dynamics are either purely real 
(outside the isentropic spinodal phase region where $v_s^2>0$) or
purely imaginary (inside the isentropic spinodal region where $v_s^2<0$).
Thus, in ideal fluid dynamics, the region of spinodal instability is bounded
by the isentropic spinodal, $v_s(\rho,T)=0$.
The introduction of viscosity adds a negative imaginary amount
to the frequency.
We then have 
$\omega_\smk^\pm\approx\zero{\omega}_k-\ihalf\lambda_{\rm visc} k^2$
to first order in $\xi\equiv\fourthird\eta+\zeta$,
where we have introduced the characteristic viscous length
$\lambda_{\rm visc}(\rho,T)\equiv\xi(\rho,T)/h_0(\rho,T)c$. But 
the inclusion of viscosity does {\em not} change the region of instability. 
In contrast, the inclusion of heat conductivity expands 
the region of instability from the isentropic spinodal 
to the isothermal spinodal, $v_T(\rho,T)=0$,
and generally increass the spinodal growth rates.

For small distortions of uniform matter at given density and temperature,
the dispersion relation yields the eigenfrequencies $w_k$ in terms of the
equation of state $p_T(\rho)$, the strength of the gradient term $C$, and the 
transport coefficients $\eta(\rho,T)$, $\zeta(\rho,T)$, and $\kappa(\rho,T)$.
We describe below what we adopt for these key functions.

%==============================================================================
\section{Transport coefficients}

The deviation of the dynamical evolution from that of an ideal fluid 
is governed by three transport coefficients:
the shear viscosity $\eta$ and the bulk viscosity $\zeta$
(which here enter only through the effective viscosity  
$\xi\equiv\fourthird\eta+\zeta$) as well as the heat conductivity $\kappa$.
Neither their magnitudes nor their dependencies on the environment 
(through $\rho$ and $T$) are very well known. 
We shall therefore employ simple parametrizations of their functional form
and introduce one adjustable overall strength parameter for each one,
thus enabling us to conveniently explore a range of physical scenarios.

%----------------------------------------------------------------------------
\subsection{Viscosity}

Using string theory methods, Kovtun {\em et al.}\ \cite{KovtunPRL94} 
made general arguments that the shear viscosity $\eta$,
in any relativistic quantum field theory
at finite temperature and zero chemical potential,
has a lower bound, $\eta\geq\hbar\sigma/4\pi$,
where $\sigma$ is the associated entropy density.

It might therefore seem natural to use $\eta=\eta_0\hbar\sigma/4\pi$,
\ie\ simply scale the minimum value by the factor $\eta_0\geq1$.
However, the entropy density $\sigma$, 
while appropriate in the context of ultrarelativistic nuclear collisions
where the medium has a high temperature and a very small net baryon density,
is not suitable in the present context where the focus is on matter 
at high baryon density and relatively modest temperature \cite{LiaoPRC81}.
A more appropriate quantity, suitable in both limits,
is the enthalpy density $h(\rho,T)\equiv p+\eps$ \cite{LiaoPRC81}.
When the net baron density $\rho$ vanishes it becomes $h=T\sigma$, whereas 
$h/c^2$ approaches the mass density at high baryon density and low temperature,
$h\approx mc^2n\gg T\sigma$, where $n$ is density of particles
and $m$ is their mass.
(For cold nuclear matter we have $n\approx\rho$, 
since the pions and anti-nucleons have negligible populations,
and hence $m=m_N$.)

Furthermore, one would 
expect the viscosity to be proportional to the interparticle spacing 
$d\equiv1/n^{1/3}$ \cite{LiaoPRC81}, 
which provides a convenient measure of the mean free path in a dense fluid.
When plasma has no net baryon density, $d$ is inversely proportional
to the temperature $T$, $\hbar c/T=4\pi c_0d$. 
The conversion constant $c_0$ is given by
\beq\label{c0}
c_0\ \equiv\ {1\over4\pi}
\left[(g_g+\threehalf g_q){\zeta(3)\over\pi^2}\right]^{1\over3}\
\approx\ 0.12779\ .
\eeq
Therefore, based on these considerations,
we shall make the following ansatz,
\beq
\label{xi}
\eta(\rho,T)\ \doteq\ \eta_0 {c_0\over c}\, d(\rho,T)\, h(\rho,T)\ ,\,\,\
	\eta_0\geq1\ ,
\eeq
which amounts to $\eta=\eta_0\hbar\sigma/4\pi$ in the baryon-free plasma,
while it becomes $\eta\approx\eta_0 c_0mcnd$ in the non-relativistic gas.
This latter expression is similar to the
familiar expression from classical transport theory 
for the shear viscosity coefficient in a dilute one-component gas, 
$\eta \approx \third m\bar{v}n\ell$,
where $\bar v$ is the mean particle speed and $\ell$ is its mean free path.

Since the bulk viscosity $\zeta$ is generally expected to be significantly
smaller than the shear viscosity $\eta$,
we shall take the effective viscosity to be
$\xi\equiv\fourthird\eta+\zeta\approx\fourthird\eta$.
It is convenient to introduce the associated characteristic length
which is proportional to the interparticle spacing,
\beq\label{lksi}
\lambda_{\rm visc}(\rho,T)\ 
\equiv\	{1\over c}{\xi(\rho,T)\over h(\rho,T)/c^2}\
=\	\fourthird \eta_0\, c_0\, d(\rho,T)\ .
\eeq

%----------------------------------------------------------------------------
\subsection{Heat conductivity}

The thermal conductivity is fundamentally related to the viscosity
because it derives from the same microscopic transport processes.
In a dilute classical gas it can be expressed as
$\kappa \approx \third \bar{v}\ell c_v$,
where $\bar v$ is the mean particle speed and
$c_v\equiv\del_T\eps_T(\rho)$ is the heat capacity
(equal to $\threehalf n$ for a dilute classical gas).
Since $\bar{p}=m\bar{v}$ and $h\approx mc^2n$,
we see that $\kappa/\eta\approx c_v/(h/c^2)$.
We therefore make the following ansatz,
\beq\label{kappa}
\kappa(\rho,T)\ 
\doteq\ \kappa_0\, c_0\, c\, d(\rho,T)\, c_v(\rho,T)\ ,\,\,\ \kappa_0\geq1\ ,
\eeq
with $c_0$ as given in Eq.\ (\ref{c0}) and assuming that the overall
strength factor $\kappa_0$ should be at least unity.
Although the relation $\kappa/\eta\approx c_v/(h/c^2)$ would suggest that
the two normalization constants should be similar, $\kappa_0\approx \eta_0$,
we prefer to leave them separately adjustable in order to make it possible
to explore the effects of the two distinct types of dissipation.
The characteristic length scale associated with heat conduction is then
\beq\label{lkappa}
\lambda_{\rm heat}(\rho,T)\ 
\equiv\	{1\over c}{\kappa(\rho,T)\over c_v(\rho,T)}\
=\	\kappa_0\, c_0\, d(\rho,T)\ .
\eeq

While the approximate expressions for the transport coefficients,
Eqs.\  (\ref{xi}) and (\ref{kappa}), should not be expected to be accurate, 
they will serve well for our present purpose of exploring the effect of the
dissipative mechanisms on the spinodal decomposition,
since they enable us to easily control the overall dissipative effects.

%----------------------------------------------------------------------------
\subsection{Growth rates}

The spinodal isothermal and isentropic boundaries 
determined by $v_T=0$ and $v_s=0$, respectively,
pertain to the thermodynamic limit of very long wave lengths, $k\to0$.
When the wave number $k$ is increased, 
the region of instability steadily shrinks
as the gradient term gives an ever larger contribution to the pressure.
Thus, 
at a given temperature $T$, the lower spinodal boundary density $\rho_A(k;T)$
increases steadily with $k$,
 while the upper spinodal boundary density $\rho_B(k;T)$ decreases steadily.
For a fixed value of $k$, a contour plot of the growth rate $\gamma_k$ 
in the $(\rho,T)$ phase plane therefore exhibits a ridge between
those two boundaries.  
The height of of the ridge decreases steadily as $T$ is increased, 
first rather gently  due to the dominance of the fermions at low temperatures.
The local value of the maximum wave number for which instabilities exist,
$k_{\rm max}(\rho,T)$, will have a similar appearance.

\begin{figure}          %       -----------------------------------------
\includegraphics[angle=0,width=3.1in]{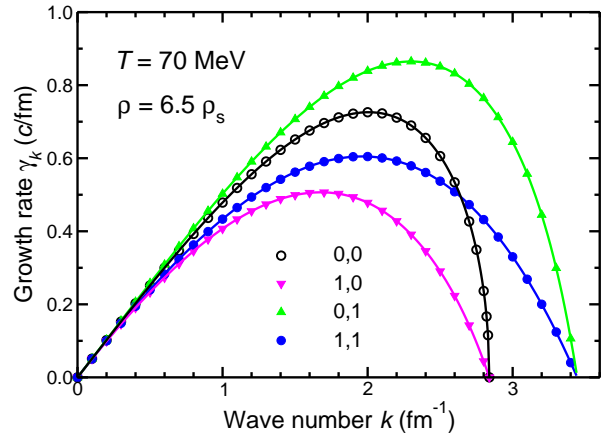}	%	dispersion relation
\caption{The growth rate $\gamma_k(\rho,T)$, as a function of the
wave number $k$, calculated with finite-range fluid dynamics
at $\rho=6.5\,\rho_s$ and $T=70\,\MeV$ 
for four different combinations of dissipation:
no dissipation ($\eta_0\!=\!0,\kappa_0\!=\!0$);
minimal viscosity but no heat conduction ($\eta_0=\!1,\kappa_0=\!0$);
no viscosity but minimal heat conduction ($\eta_0=\!0,\kappa_0=\!1$); both
minimal viscosity and minimal heat conduction ($\eta_0\!=\!1,\kappa_0=\!1$).
}\label{f:gamma}
\end{figure}            %       -----------------------------------------

We now illustrate, in Fig.\ \ref{f:gamma}, the resulting dispersion relations
for thermodynamic scenarios relevant to the present study.
Selecting a phase point in the central region of the phase coexistence region
where both the isothermal and the isentropic sound velocities are imaginary,
$\rho=6.5\rho_s$ and $T=70\,\MeV$ (see Fig.\ \ref{f:T-c}),
we consider the growth rate $\gamma$ as a function of the wave number $k$ 
of the density undulation being amplified.
The non-dissipative treatment with ideal finite-range fluid dynamics
provides a convenient reference result.
It yields a fastest growth time of about $t_0\approx1.38\,\fm/c$
which occurs for wave numbers near $k_0\approx2.0\,\fm^{-1}$,
corresponding to an optimal wave length of $\ell_0=2\pi/k_0\approx3.14\,\fm$.

Relative to this reference, the inclusion of viscosity slows the growth
but does {\em not} change the domain of instability
which is still delineated by the vanishing of the isentropic sound speed $v_s$.
We see that the inclusion of a minimal amount of viscosity ($\eta_0=1$)
leads to a significant reduction in $\gamma$ and also shifts
the optimal length scale towards larger values.

On the other hand, relative to the ideal scenario,
the inclusion of heat conductivity enlarges the domain of instability,
the boundary being now determined by the vanishing of the
isothermal sound speed $v_T$.
Thus, generally, the inclusion of heat conductivity increases the growth rates,
particularly at the high end of the unstable $k$ range.

While the inclusion of both minimal viscosity and minimal heat conduction
necessarily enlarges the unstable $k$ range, it does somewhat reduce the
fastest growth rates.  However, 
it hardly affects the scale of the fastest-growing modes, $k_{\rm max}$.
As the strengths of the dissipative terms are further increased,
the growth rate $\gamma_k$ decreases steadily and, at the same time,
the maximum in the dispersion relation moves gradually downwards in $k$.

\begin{figure}          %       -----------------------------------------
\includegraphics[angle=0,width=3.1in]{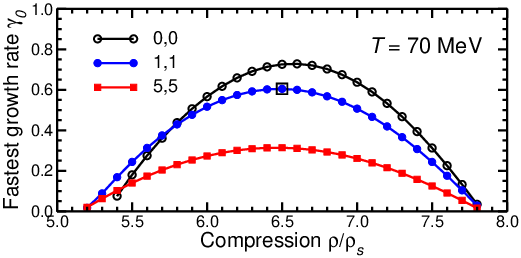}	%	max growth rate
\includegraphics[angle=0,width=3.1in]{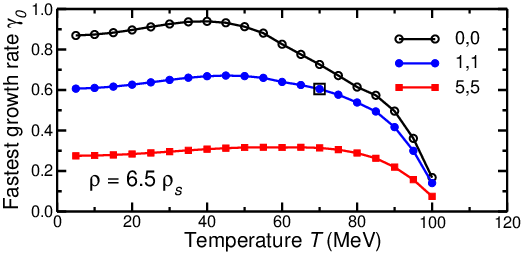}	%	max growth rate
\caption{The maximum growth rate $\gamma_0$ ($c$/fm)
shown at various compressions for fixed $T=70\,\MeV$ ({\em upper panel}) and
at various temperatures for fixed compression $\rho/\rho_s=6.5$ 
({\em lower panel}),
for three different assumptions about the dissipation, namely
no dissipation: $(\eta_0,\kappa_0)=(0,0)$;
minimal dissipation: $(1,1)$, and strong dissipation: $(5,5)$.
The open square corresponds to the phase point at which the dispersion relation
shown in Fig.\ \ref{f:gamma} was obtained (using minimal dissipation).
}\label{f:gmax}
\end{figure}            %       -----------------------------------------

These features are present throughout the unstable region
of the $(\rho,T)$ phase plane.
In order to obtain an impression of how the growth rates change
with $\rho$ and $T$,
we show in Fig.\ \ref{f:gmax} the fastest growth rate $\gamma_0$
as we move away from the phase point explored in Fig.\ \ref{f:gamma},
either at constant $T$ ({\em upper panel}) 
or constant $\rho$ ({\em lower panel}), as indicated in Fig.\ \ref{f:T-c}.
In these illustrations we show the ideal scenario where no dissipation
is present ($\eta_0, \kappa_0=0$),
the scenario with minimal dissipation ($\eta_0, \kappa_0=1$),
and a scenario with five times stronger dissipation ($\eta_0, \kappa_0=5$),
thus covering the range suggested by the RHIC data
\cite{Heinz:2009xj,Teaney:2009qa,SongJPG36};
the currently favored estimate is $\eta_0\approx1$--$3$.
Since, as noted above, the heat conductivity is fundamentally related
to the shear viscosity, it should be expected that their strengths vary
in approximate unison; the effect of varying their relative strengths
may be judged from the results shown in Fig.\ \ref{f:gamma}.

As expected from the discussion in the beginning of this section,
the fastest growth at a given temperature 
occurs about midway between the corresponding spinodal boundaries.
Furthermore, the temperature generally reduces the growth rates,
though only weakly at small $T$.
The curves in the lower panel of Fig.\ \ref{f:gmax} do not exhibit 
a monotonic decrease because the selected path along a constant density 
does not follow the ridgeline:
the maxima in the curves shown in the upper panel shift steadily 
towards smaller densities as $T$ is increased,
as would be expected from the  general appearance of the phase diagram.

Finally,
we note that a five-fold increase in the dissipation above the minimal value,
{\em i.e.}\ changing $(\eta_0,\kappa_0)$ from $(1,1)$ to $(5,5)$,
leads to a reduction in the growth rates $\gamma_k$ 
by only about a factor of two.

%==============================================================================
\section{Dynamical evolution}

After the above preparations, we are now in a position to address
the dynamical evolution of the unstable collective modes
in the spinodal region of the phase diagram.

%------------------------------------------------------------------------------
\subsection{Dynamical phase trajectories}

\begin{figure}          %       -----------------------------------------
\includegraphics[angle=0,width=3.1in]{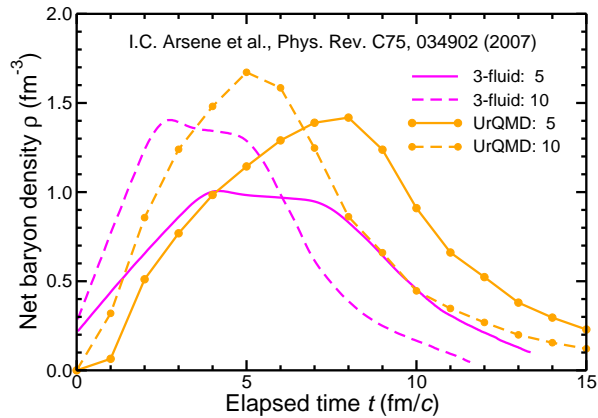}	%	rho-t
\caption{The time evolution of the net baryon density, $\rho(t)$,
at the center of a head-on gold-gold collision 
for bombarding energies of 5 and 10 $\GeV/A$,
as calculated with the 3-fluid \cite{3-fluid}
and the UrQMD \cite{UrQMD} models 
(from Ref.\ \cite{ArsenePRC75}).
}\label{f:rho-time}
\end{figure}

We first specify dynamical phase trajectories, $(\rho(t),T(t))$,
that are representative of the bulk matter in the collision zone,
making use of the results presented in Ref.\ \cite{ArsenePRC75}.
In that work, a number of different dynamical models were used to extract the
time evolution of the net baryon density, $\rho(t)$, and the energy density,
$\eps(t)$, in the center of a head-on gold-gold collision
for the range of collision energies anticipated at FAIR.
The resulting dynamical trajectories in the $(\rho,\eps)$ phase plane 
were remarkably independent of the specific model, in large part
probably due to the robust nature of the mechanical densities which are
subject to conservation laws, in contrast to the corresponding thermodynamical
variables $(\mu,T)$.
Nevertheless, there were significant variations in the detailed behavior.
This is illustrated in Fig.\ \ref{f:rho-time} 
which shows density evolutions $\rho(t)$
obtained with the 3-fluid model \cite{3-fluid} and UrQMD \cite{UrQMD}
at beam kinetic energies of $5$ and $10$ $\GeV/A$.
(Note that these beams are bombarded onto stationary targets,
as will be done at FAIR; the same collision energies can be obtained
in a collider configuration by using total beam energies
of $\approx1.8$ and $\approx2.4$ $\GeV/A$, respectively.)
We utilize these results to construct the dynamical
phase trajectories considered.

            %       -----------------------------------------

\begin{figure}          %       -----------------------------------------
\includegraphics[angle=0,width=3.1in]{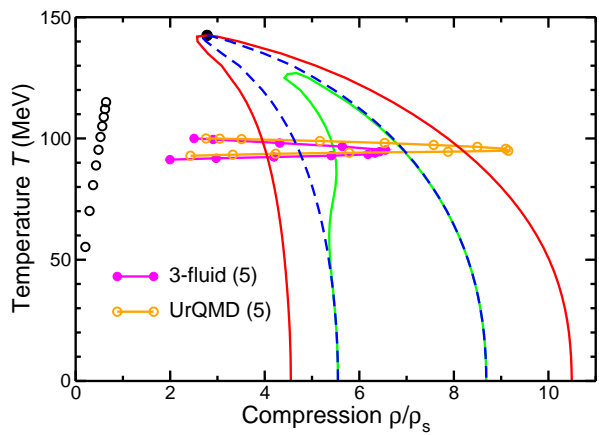}	%	rho-t
\caption{Dynamical phase trajectories based on the 3-fluid and UrQMD 
density evolutions (shown in Fig.\ \ref{f:rho-time})
obtained for 5~GeV/$A$ in Ref.\ \cite{ArsenePRC75};
the associated time-dependent growth rates $\gamma_\smk(t)$
are illustrated in Fig.\ \ref{f:gammat}.
The symbols along the trajectories are equidistant in time
with $\Delta t=1\,\fm/c$, while the open dots on the left
indicate the freezeout locations for bombarding energies of 
$E=1,\dots,10\,\GeV/A$ obtained from fits to experimental data 
as discussed in Ref.\ \cite{RandrupPRC74}.
}\label{f:traj}
\end{figure}            %       -----------------------------------------

\begin{figure}          %       -----------------------------------------
\includegraphics[angle=0,width=3.1in]{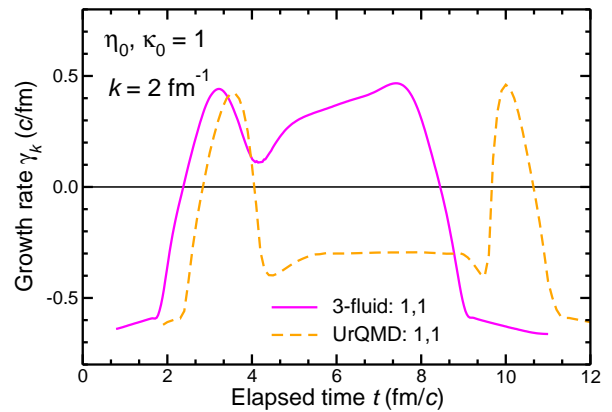}	%	gamma(t)
\caption{The spinodal growth rate $\gamma_\smk(t)={\rm Re}[\omega_\smk(t)]$ 
for modes with $k=2\,\fm^{-1}$, calculated with minimal dissipation 
along the two dynamical $(\rho,T)$ phase trajectories 
shown in Fig.\ \ref{f:traj}.}\label{f:gammat}
\end{figure}            %       -----------------------------------------

As expected, the degree of amplification achieved
depends strongly on the length of time spent in the phase region
of spinodal instability.
In order to elucidate this key feature, we consider in some detail
the phase trajectories depicted on Fig.\ \ref{f:traj}.
The density evolutions $\rho(t)$ are those calculated with
the two models for $5~\GeV/A$, 
while the temperature evolutions $T(t)$ are (somewhat arbitrarily) prescribed
with an eye towards the freeze-out conditions extracted from data
\cite{RandrupPRC74}.
The essential difference between these two trajectories is that
UrQMD yields compressions that reach all the way 
into the deconfined phase region, 
whereas the maximum compression achieved with the 3-fluid model 
lies inside the unstable region.
Such a situation would be obtained with UrQMD as well
at a somewhat lower collision energy (3-4~$\GeV/A$).

When the turning point of the phase trajectory lies inside the unstable region
the modes are exposed to the spinodal amplification for a longer time and,
presumably, the resulting degree of amplification would be maximized.
This key fact is illustrated in Fig.\ \ref{f:gammat}.
For the penetrating (UrQMD) trajectory, the modes are exposed to
amplification only during the two relatively brief periods 
when the phase point is traversing the spinodal region
(and the amplification gained during the compressional traverse
is largely lost during the high-density stage due to the equilibration process
so only the amplification received during the expansion traverse is relevant).
By contrast, the more optimal (3-fluid) trajectory
provides amplification for a sustained period of time,
thus making a phase separation more likley to occur.

While it is yet difficult
to make specific predictions about the optimal collision energy,
it seems evident that such an energy range exists, 
since the location of the $(\rho,T)$ phase point 
associated with the maximum compression (the turning point)
moves steadily downwards as the collision energy is lowered.
The experimentally known freeze-out temperatures
(indicated on the left in Fig.\ \ref{f:traj}) provide a lower bound
on the phase region that could be accessed by nuclear collisions.
The turning point is therefore likely to traverse the unstable phase region
at fairly high temperatures where it is relatively narrow.
Consequently, the optimal range of collision energy is most likely 
not so wide.
Because of this generic expectation, we shall take the above phase
trajectories as being reasonably representative of what may happen
in the bulk region of an actual collision.

%------------------------------------------------------------------------------
\subsection{Dissipative collective dynamics}

As the bulk of the evolving system traverses the unstable phase region,
its density fluctuations may be amplified.
In order to make quantitative estimates of this effect,
we employ the method developed in Ref.\ \cite{ColonnaNPA567}
for the evolution of collective modes subject to a dissipative coupling
to the environment, 
\ie\ to the (many more) non-collective modes in the system.

The treatment considers an ensemble of $\cal N$ macroscopically 
similar systems $\{n\}$ 
that have all been prepared with the same average density $\rho_0$
relative to which individual perturbations are introduced,
\beq
\rho^{(n)}(\bfr)\ =\ \rho_0+\delta\rho^{(n)}(\bfr)\ 
=\ \rho_0+ \sum_{\smk\neq0}\rho_\smk^{(n)} \rme^{i\smk\cdot\smr} .
\eeq
Generally, such as during the expansion stage of a nuclear collision,
the bulk density is time dependent.
We shall assume that the effect of this overall evolution 
may be taken into account approximately
by using the corresponding time-dependent transport coefficients 
in the formulas below.
We should then think of the mode index $\bfk$ as a mode number $\bold{K}$ 
rather than a wave number,
since the expansion primarily stretches the modes without 
introducing much mode mixing \cite{ColonnaPRC51}.
In this way we may obtain the spatial size of the resulting fluctuations
by scaling the initial wave length with the linear expansion factor,
$[\rho_0(t_i)/\rho_0(t_f)]^{D/3}$,
where $D$ is the effective dimensionality of the expansion.

The primary object of study is the associated spatial correlation function,
$\sigma(\bfr_{12}) = \prec \delta\rho(\bfr_1)\,\delta\rho(\bfr_2)^*\succ$,
where
\beq
\prec \delta\rho(\bfr_1)\,\delta\rho(\bfr_2)^*\succ\ \equiv
{1\over\cal N}\sum_n \delta\rho^{(n)}(\bfr_1)\ \delta\rho^{(n)}(\bfr_2)^*\ 
\eeq
is the ensemble average and $\bfr_{12}\equiv\bfr_1-\bfr_2$.
The Fourier transform of $\sigma(\bfr)$ provides a convenient measure 
of the degree of density fluctuation at a given scale,
since it is the ensemble average of $|\rho_\smk|^2$,
\beq\label{sigma2k}
\sigma_\smk^2\ =\ \prec|\rho_\smk|^2\succ\
=\	\int{d\bfr\over V}\, \rme^{-i\smk\cdot\smr} \sigma(\bfr)\ .
\eeq

We are generally interested in the evolution of the collective modes
in the system.
Adopting the treatment of Ref.\ \cite{ColonnaNPA567},
we assume that the amplitude $\rho_\smk$ of a given collective mode $\bfk$
is governed by an equation of motion having the form
\beq
{d\over dt}\rho_\smk(t)\ =\ -i\omega_\smk(t) \rho_\smk(t) +B_\smk(t)\ .
\eeq
The complex eigenfrequencies $\omega_\smk=\epsilon_\smk+i\gamma_\smk$
are determined by the dispersion equation derived above,
while the Brownian term $B_\smk$ describes the residual coupling between
the collective mode and the reservoir, 
{\em i.e.}\ the non-collective modes of the system.
It is fluctuating in nature and is assumed to be Markovian,
\beq
\prec B_\smk(t)\,B_{\smk'}(t')^*\succ\ 
	=\ 2{\cal D}_{\smk\smk'}(t)\,\delta(t-t')\ .
\eeq

The equal-time correlation coefficient for two collective modes,
$\sigma_{\smk\smk'}(t)\equiv\prec\rho_\smk(t)\rho_{\smk'}(t)^*\succ$,
is then given by \cite{ColonnaNPA567},
\beq
\sigma_{\smk\smk'}(t) =
	\sigma_{\smk\smk'}(t_i)\,\rme^{-i\omega_{\smk\smk'}t}\!
 	+2\!\int_{t_i}^t\! dt'\, {\cal D}_{\smk\smk'}(t')\,
	\rme^{i\omega_{\smk\smk'}(t'-t)} ,
\eeq
where $\omega_{\smk\smk'}\equiv\omega_\smk-\omega_{\smk'}^*$.
It is readily seen that it satisfies the following differential equation,
\beq\label{lalime}
{d\over dt}\sigma_{\smk\smk'}(t)\ =\ 
-i\omega_{\smk\smk'}(t)\sigma_{\smk\smk'}(t)+2{\cal D}_{\smk\smk'}(t)\ ,
\eeq
which was dubbed the {\em Lalime equation} in Ref.\ \cite{ColonnaNPA567}.

For our present studies, we are particularly interested in the time evolution
of the diagonal components of the covariance matrix, 
$\sigma_{\smk\smk}\!=\!\sigma_\smk^2$,
which are equal to the fluctuation coefficients %$\sigma_\smk^2$ 
introduced in Eq.\ (\ref{sigma2k}). Then
\beq
\sigma_\smk^2(t)\ =\
\left[\sigma^2_\smk(t_i)+\int_{t_i}^t 2{\cal D}_\smk(t')\,
	\rme^{-2\Gamma_\smk(t')}dt'\right]\rme^{2\Gamma_\smk(t)}\ ,
\eeq
where ${\cal D}_\smk\equiv{\cal D}_{\smk\smk}$
and we have introduced the following {\em amplification coefficient},
\beq\label{Gamma}
\Gamma_\smk(t)\ \equiv\ \int_{t_i}^t{\rm Im}[\omega_\smk(t')]\,dt'\
=\ \int_{t_i}^t\gamma_\smk(t')\,dt'\ . 
\eeq

If the environment is stationary,
then $\gamma_\smk$ and ${\cal D}_\smk$ remain constant in time,
so $\Gamma_\smk(t)=\gamma_\smk(t-t_i)$, 
and we obtain a simple exponential time evolution,
\beq
\sigma_\smk^2(t)\ =\ {{\cal D}_\smk\over\gamma_\smk}
\left[\rme^{2\gamma_\tnk(t-t_i)}-1\right]
	+\sigma^2_\smk(t_i)\,\rme^{2\gamma_\tnk(t-t_i)}\ .
\eeq
When $\gamma_\smk$ is negative, as is normally the case,
the mode is stable and relaxes
exponentially towards its equilibrium value $\tilde{\sigma}^2_\smk$,
$\sigma^2_\smk(t)\to-{\cal D}_\smk/\gamma_\smk$,
the associated relaxation time being $t_\smk=1/\gamma_\smk$.
More generally, within the stable regime, $\sigma^2_\smk(t)$
 will seek to relax towards its instantaneous equilibrium value 
$\tilde{\sigma}^2_\smk(t)=-{\cal D}_\smk(t)/\gamma_\smk(t)$,
in accordance with the Einstein relation.
In the opposite case, when $\gamma_\smk$ is positive, the mode is unstable
and the fluctuations exhibit an exponential growth,
with both the original mode fluctuations $\sigma^2_\smk(t_i)$
and the noise ${\cal D}_\smk/\gamma_\smk$ being amplified.

The diffusion coefficients ${\cal D}_\smk$ 
represent the coupling of the collective modes to the residual system. 
They thus play two important roles 
in the dynamical evolution of the density fluctuations.
In the stable regime (as mentioned above) 
the coupling determines the relaxation times $t_\smk$
 for the relaxation of the fluctuation coefficients $\sigma^2_\smk$ 
towards their appropriate equilibrium values $\tilde{\sigma}^2_\smk$,
while in the unstable regime it continually produces additional fluctuations
that are also amplified.
Thus, even in the absense of initial fluctuations, 
the coupling of the collective mode to the residual system 
will continually create fluctuations
that will subsequently become amplified or damped,
as governed by the Lalime equation (\ref{lalime}).
In particular, the diffusion coefficients enable the stable modes
to continually adjust their fluctuations
as the equilibrium variances evolve.

\begin{figure}          %       -----------------------------------------
\includegraphics[angle=0,width=3.1in]{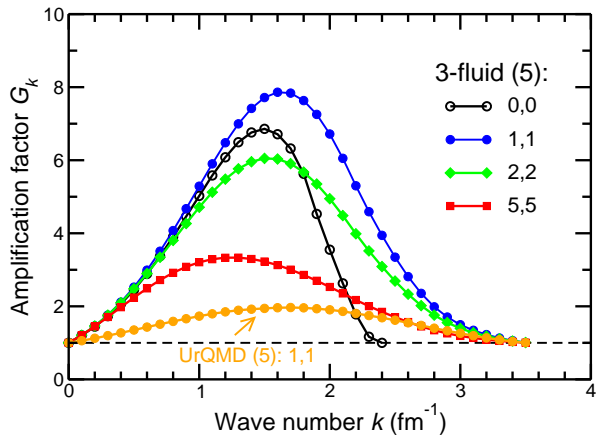}	%	Gk
\caption{As a function of the wave number $k$ is shown
the amplification factor $G_k$ (see Eq.\ (\ref{Gk}))
resulting from motion
along the 3-fluid phase trajectory displayed in Fig.\ \ref{f:traj}
for various degrees of dissipation
(indicated by the values of $\eta_0$ and $\kappa_0$).
Also shown is the result for the UrQMD trajectory in Fig.\ \ref{f:traj}
using minimal dissipation, \ie\ $(\eta_0,\kappa_0)=(1,1)$.
}\label{f:Gk}
\end{figure}            %       -----------------------------------------

%------------------------------------------------------------------------------
\subsection{Onset of phase separation}

The amplification coefficient $\Gamma_\smk$ defined in Eq.\ (\ref{Gamma})
is obtained by integrating $\gamma_\smk\equiv{\rm Im}(\omega_\smk)$ 
over the entire time interval considered.  
When there is no dissipation, $\omega_\smk^2$ is always real 
so $\gamma_\smk$ vanishes outside the unstable region.
But in the presence of dissipation, $\gamma_\smk$ is negative outside
the unstable region, and increasingly so the stronger the dissipation.
As a result, the local relaxation time is reduced which in turn
ensures that the fluctuations remain close to their equilibrium value 
until shortly before the trajectory enters the unstable region.
However, by the same token, any amplification acquired while the trajectory
is inside the unstable region is correspondingly quickly lost 
after the trajectory has reentered the normal region and again equilibrates.
In fact, the dissipative length tends to be larger at late times,
due to the increased particle spacing in the more dilute system,
thus shortening the relaxation time.

Our present study aims to clarify the prospects for the spinodal amplification
to become sufficiently large to cause a phase-separating clumping 
of the system.
Since our treatment is perturbative, it is only valid as long as
the fluctuations remain relatively small.
The occurrence of large fluctuations in the calculation 
should then be taken as a signal that clumping is likely.
It should be recalled that the bulk of the system is thermodynamcially
unstable 
so even a modest degree of fluctuation may cause a catastrophic breakup.
The prospects for this to occur are enhanced by the fact that fluctuations
created in the mechanically unstable (spinodal) region may become futher
amplified during the traverse of the adjacent metastable region,
just as impurities introduced in this region may trigger condensation.

Thus the key question is how much amplification may occur during the
unstable era.
To elucidate this issue in a quantitative manner,
we extract the following {\em amplification factor},
\beq\label{Gk}
G_\smk\ \equiv\ \exp(\int_>\!\gamma_\smk(t\,)dt)\ ,
\eeq
where the integral is only over those times during which the mode is unstable,
$\gamma_\smk>0$.

Figure \ref{f:Gk} shows values of $G_k$ obtained
for the entire range of wave numbers,
as obtained with various degrees of dissipation.
While we concentrate on the phase trajectory that reaches its maximum
compression inside the spinodal region 
(the one based on the 3-fluid model results for $5~\GeV/A$),
we also show the result of a more penetrating trajctory
to bring out the importance of tuning the collision energy for optimal effect.

In addition to the result of ideal fluid dynamics,
we show results for various degrees of dissipation,
ranging from minimal, \ie\ $(\eta_0,\kappa_0)=(1,1)$, to five times that.
While viscosity generally slows the evolution,
thus also supressing the growth of instabilities,
heat conductivity generally increases the growth rate.
The combined effect of introducing small amounts of dissipation 
then tends to enhance the amplification.
Thus the resulting degree of non-uniformity is fairly robust 
against moderate changes in the dissipation strength and,
consequently, our conclusions do not appear to be sensitive to
the specific parametrizations of the transport coefficients.

The results displayed in Fig.\ \ref{f:Gk} 
bring out the characteristic feature of spinodal instability,
namely that the amplification mechanism favors certain length scales.
We note that the two-point correlation coefficient $\sigma_k^2$
is proportional to $G_k^2$ and thus exhibits a stronger peaking.
More generally, since the $N$-point correlation is proportional to $G_k^N$,
the spinodal effect manifests itself progressively stronger
in the higher-order correlations.

We see that $G_k$ is peaked around $k_0\approx1.6$--$1.3\,\fm^{-1}$,
depending on the degree of dissipation,
which corresponds to wave lengths $\ell_0=2\pi/k_0\approx3.9$--$4.8\,\fm$.
If, at a temperature of 80--100~MeV,
a uniform system situated at the lower edge of the spinodal region
(hence having a density $\rho_0\approx5\rho_s$)
transforms itself into plasma drops embedded in a hadron gas,
with each subsystem having the corresponding coexistence density,
$\rho_1(T)\approx4\rho_s$ and $\rho_2(T)\approx8\rho_s$,
then we may expect the drop radius to be
$R_{\rm drop}\approx\half\ell_0[(\rho_0-\rho_1)/(\rho_2-\rho_0)]^{1/3}
\approx 1.2$--$1.5\,\fm$.
Each such drop would then have a baryon number of 10--18.
Since this is only a relatively small fraction of the collision system
one would expect that several such drops would be formed
during the phase separation.
On the other hand, a drop of such a size 
is sufficiently large to be regarded as a macroscopic source 
that will undergo statistical hadronization.

%==============================================================================
\section{Concluding remarks}

In the exploration of the phase diagram of strongly interacting matter
by means of nuclear collisions,
the mechanism of spinodal phase decomposition might give rise
to unique signals of the first-order phase transition.
For the planning of possible experimental campaigns to search for
spinodal phase separation, it is useful to have estimates of which
bombarding energies are expected be be optimal for producing the effect.
The analysis presented above suggests that these bombarding energies lie
at the lower end of the anticipated FAIR range.
It is also well within the energy region proposed for NICA
but appears to be too low for experiments at RHIC to be feasible.

However, 
when making a quantitative prediction of the optimal collision energy range,
it should be kept in mind that the calculated dynamical phase trajectories
\cite{ArsenePRC75},
on the basis of which we made our estimates,
extracted the conditions right at the center of the collision zone.
Therefore they most likely provide an upper limit on the compression
and it must be exptected that the average compression and excitation 
over an extended volume will be smaller than 
those shown in Figs.\ \ref{f:rho-time} and \ref{f:traj}.
Consequently, the optimal bombarding energy is probably somewhat higher
than what the above idealized analysis would suggest.
For example, it may well be that the $5\,\GeV/A$ UrQMD simulation,
whose maximum compression overshoots the spinodal region,
would be quite suitable for spinodal clumping because the average degree
of compression over an extended region favors phase separation.

It should also be recognized that the spinodal phase region is
surrounded by a thermodynamically metastable region through which the
phase trajectory has to pass after the spinodal mechanism has acted.
During this traverse, sufficiently significant deviations from uniformity
will be further amplified and phase separation is thus more likely to occur
than suggested by just the perturbative treatment employed here.
In fact, the spinodal amplification of the preexisting fluctuations
may be regarded as merely providing seeds for a subsequent non-linear
breakup evolution in the metastable region.

We note the analogy with the search for spinodal fragmentation 
\cite{PhysRep389}
as a signal of the nuclear liquid-gas phase transition \cite{BorderiePRL86}.
The collision energy had to be carefully adjusted to producing the
phenomenon: a too high energy would yield an explosive vaporization,
while a too low energy would cause the two nuclei to fuse 
(and subsequently deexcite by light-particle emission).  
But in the optimal energy range,
the bulk of the combined system would first compress (to 2--3 times normal)
and then start to expand; however, the expansion would tend to stall,
leaving the system in a dilute (and nearly spherical) configuration
for a sufficient length of time (several times the typical growth time)
to cause the most unstable modes to grow dominant, thereby leading the
system towards a breakup into nearly equal-sized intermediate-mass fragments.
Our present analysis suggests that collisions at suitably tuned relativistic
energies would also cause the bulk of the system to spend several growth times
inside the unstable phase region 
and thus enable the spinodal formation of plasma drops.

In the liquid-gas case the production of equal-size nuclear fragments
in each event provided a simple and unambigous signal 
of the spinodal breakup mechanism
and thus for the existence of a first-order transition. 
The confinement transition is inherently more difficult to investigate
experimentally because any plasma drops that may have been formed
will ultimately hadronize and are thus harder to identify.
Nevertheless, the transient existence of such spatially separated
blobs of deconfined matter may be revealed by careful examination
of suitable multi-particle correlations.
While some relatively schematic studies have already been made
for the purpose of identifying such observables 
\cite{BowerPRC64,RandrupHIP22,KochPRC72}
there is a need for much more refined studies.
We hope that the present investigation, 
which suggests that spinodal phase separation might indeed occur,
will provide an incentive for such endeavors.

%---------------------------------------------------------------------------
\section*{Acknowledgements}
We wish to acknowledge helpful discussions with
V.~Koch, J.~Liao, H.C.~Song, and D.N.~Voskresensky.
This work was supported by the Director, Office of Energy Research,
Office of High Energy and Nuclear Physics,
Nuclear Physics Division of the U.S.\ Department of Energy
under Contract No.\ DE-AC02-05CH11231.

\appendix
%==============================================================================
\section{Equation of state}
\label{EoS}

The present study requires an equation of state of strongly interacting matter
that displays the expected phase structure.  
Although significant progress has been made in understanding the
thermodynamical properties of each of the phases separately,
our current understanding of the phase coexistence region is not yet
on firm ground.  We therefore employ a conceptually simple approximate
equation of state which will suffice for our present explorations.

For this purpose, we approximate the confined phase by an ideal gas of
nucleons and pions augmented by a density-dependent interaction energy,
while the deconfined phase is taken as an ideal gas of quarks and gluons
with their interactions described by a bag constant.
The desired phase structure is then generated by suitable interpolation 
between these two pure phases.

%----------------------------------------------------------------------------
\subsection{Confined phase}

The confined phase is approximated as an ideal gas of pions,
nucleons, and antinucleons,  plus an interaction term.
The total hadronic pressure is thus
\beq
p^H = p_\pi+p_N+p_{\bar N}+p_w ,
\eeq
where the contribtion from the ideal pion gas is
\beq
p_\pi(T)\ =\ -g_\pi T\int_{m_\pi}^\infty{p\epsilon d\epsilon\over2\pi^2}
	\ln[1-\rme^{-\beta\epsilon}]\ ,
\eeq
with $g_\pi=3$ and $m_\pi=140\,\MeV$, 
while the nucleons and antinucleons contribute
\beqar
p_N(T,\mu_0) &=& g_N\int_{m_N}^\infty{p\epsilon d\epsilon\over2\pi^2}
	\ln[1+\rme^{-\beta(\epsilon-\mu_0)}]\ ,\\
p_{\bar N}(T,\mu_0) &=& g_N\int_{m_N}^\infty{p\epsilon d\epsilon\over2\pi^2}
	\ln[1+\rme^{-\beta(\epsilon+\mu_0)}]\ ,
\eeqar
respectively, with $g_N=2\times2=4$ and $m_N=940\,\MeV$.
The net baryon density $\rho^H=\rho_N-\rho_{\bar N}$ then follows,
\beq
\rho^H = {\del p^H\over\del\mu_0} =
g_N \int_{m_N}^\infty{p\epsilon d\epsilon\over2\pi^2}
{\sinh\beta\mu_0\over\cosh\beta\mu_0+\cosh\beta\epsilon}\ ,
\eeq
and the entropy density is $\sigma^H=\del p^H/\del T$.
Finally, the contribution from the interaction energy density $w(\rho)$ is
$p_w(\rho)=\rho\del_\rho w(\rho)-w(\rho)$
and the parameter $\mu_0$ is related to the chemical potential $\mu$ by
$\mu=\mu_0+\del_\rho w$.

\begin{figure}          %       -----------------------------------------
\includegraphics[angle=0,width=3.1in]{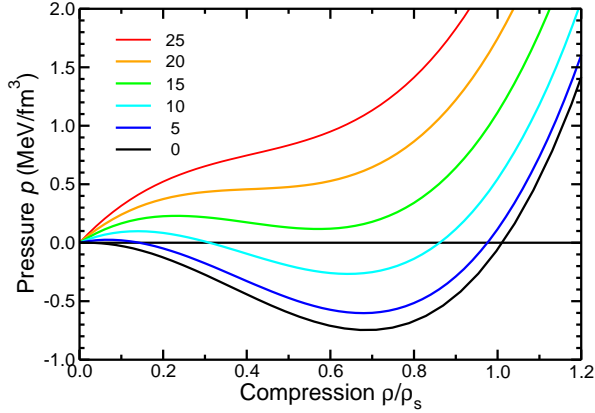}	%	Nuclear EoS
\caption{The equation of state, $p_T(\rho)$,
for the $(\rho,T)$ region relevant for ordinary nuclear matter:
the dependence of the pressure on the compression %$\rho/\rho_s$
for specified $T$ (shown in MeV).
}\label{f:pw}
\end{figure}

The interaction energy density $w(\rho)$ has the form
\beq
w(\rho)\ =\ \left[-A\left({\rho\over\rho_s}\right)^\alpha
		  +B\left({\rho\over\rho_s}\right)^\beta\right]\rho\ ,
\eeq
where we use $\alpha=1$ and $\beta=2$.
The strength coefficients $A$ and $B$ are then adjusted
so that nuclear matter saturates at $\rho_s\!=\!0.153\,\fm^{-3}$
and the associated compression modulus is
$K=K_N+K_w=300\,{\rm MeV}$, 
where $K_N=-\mbox{$6\over5$}E_F\approx-43\,{\rm MeV}$
is the contribution from the Fermi motion of the nucleons
(which is {\em negative}).
The binding energy of nuclear matter is then also roughly reproduced.
The resulting equation of state for nuclear matter is shown
 in Fig.\ \ref{f:pw}.

            %       -----------------------------------------

%----------------------------------------------------------------------------
\subsection{Deconfined phase}

The deconfined phase is taken as an ideal gas of massless gluons 
and light quarks with a standard bag constant,
\beq
p^Q\ =\ p_g+p_q+p_{\bar{q}}-B\ ,
\eeq
where $p_g=g_g(\pi^2/90)T^4$ with $g_g=2\times8=16$ is the gluon pressure
while the quarks and antiquarks contribute
\beq
p_q+p_{\bar{q}} = g_q
\left[{7\pi^2\over360}T^4+{1\over12}\mu_q^2T^2+{1\over24\pi^2}\mu_q^4\right]\ ,
\eeq
with $g_q=2\times3\times2=12$ and $\mu_q=\third\mu$.
The net baryon density in the plasma is then
\beq
\rho^Q\ =\ {\del p^Q\over\del\mu}\ =\ {2\over9}\mu T^2+{2\over81\pi^2}\mu^3\ ,
\eeq
while the entropy density is
\beq
\sigma^Q\ =\ {\del p^Q\over\del T}\ =\ 
{74\over45}\pi^2T^3+{2\over9}\mu^2T\ .
\eeq
For the bag constant we use $B=300\,\MeV/\fm^3$.

%----------------------------------------------------------------------------
\subsection{Interpolation}

\begin{figure}          %       -----------------------------------------
\includegraphics[angle=0,width=3.1in]{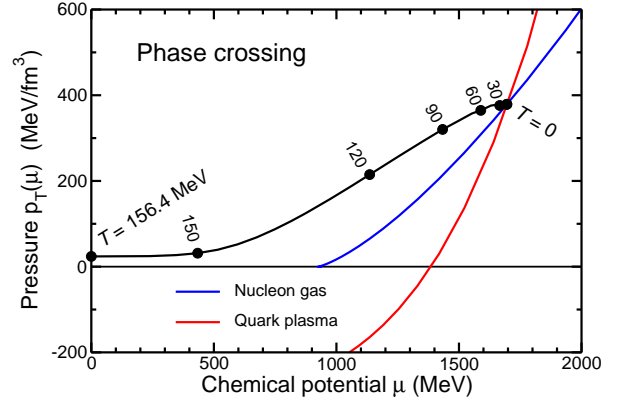}	%	phase crossing
\caption{Phase crossing: The pressures in the two idealized phases 
are shown as functions of the chemical potential $\mu$ for $T=0$;
the systems are in mutual thermodynamic equilibrium at the $\mu$ value 
for which the two curves cross.
The crossing points obtained by the same procedure for $T>0$ are connected 
by the solid curve which terminates at $T_{\rm max}\approx156.4\,\MeV$.
}\label{f:x-mu}
\end{figure}

At zero temperature and zero chemical potential,
the pressure of the nucleon gas vanishes
while that of the quark gas is equal to $-B$.
The confined phase is then the thermodynamically favored one.
However, as the chemical potential is raised,
the plasma pressure increases faster than the hadronic pressure,
so the two curves, $p^H(T=0,\mu)$ and $p^Q(T=0,\mu)$ cross
at a certain value of $\mu$, 
above which the deconfined phase is favored,
as illustrated in Fig.\ \ref{f:x-mu}.
This phase crossing procedure can be repeated 
for any temperature up to $T_{\rm max}$ 
and the resulting crossing points are included in Fig.\ \ref{f:x-mu}.

            %       -----------------------------------------

\begin{figure}          %       -----------------------------------------
\includegraphics[angle=0,width=3.1in]{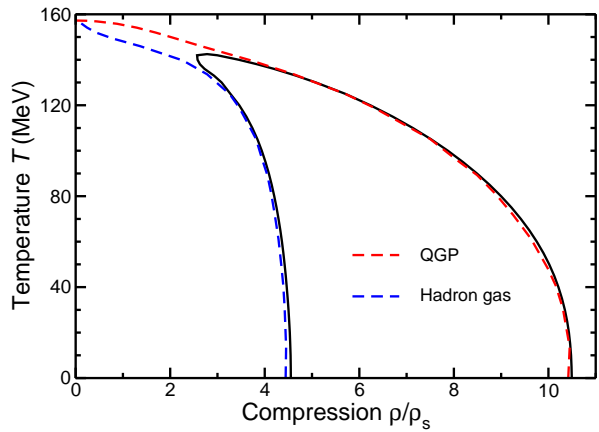}	%	phase crossing
\caption{Phase crossing:
For the range of temperatures where the two idealized phases coexists,
$0\leq T\leq T_{\rm max}\approx156.4$~MeV,
the associated coexistence densities are shown (dashed curves).
The solid curves show the corresponding coexistence densities
for the spline-based  equation of state.
}\label{f:rho-Tx}
\end{figure}            %       -----------------------------------------

For the discussion of phase coexistence, it is convenient to work in the
canonical representation where the temperature is specified.
Then the condition of phase coexistence, 
{\em i.e.}\ same temperature, chemical potential and pressure 
at two different densities $\rho_1$ and $\rho_2$,
amounts to the condition that $f_T(\rho)$, 
the free energy density as a function of density,
have common tangents.
This is readily seen since $\mu_T(\rho)=\del_\rho f_T(\rho)$
implies that the two chemical potential are then equal, $\mu_1=\mu_2$,
and since the tangent at $\rho_i$ is given by 
$t_i(\rho)=f_i(\rho)+(\rho-\rho_i)f_i'(\rho)$ 
the fact that $t_1(\rho)=t_2(\rho)$ immediately implies $p_1=p_2$.

\begin{figure}          %       -----------------------------------------
\includegraphics[angle=0,width=3.1in]{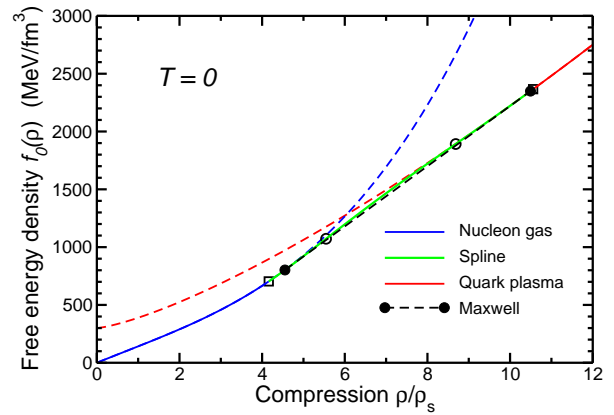}	%	phase crossing
\caption{The free energy density at zero temperature, $f_{T=0}(\rho)$,
as a function of the degree of compression $\rho/\rho_c$.
The two individual phases $f_0^H(\rho)$ and $f_0^Q(\rho)$
are shown together with the connection between them, $\tilde{f}_0(\rho)$,
obtained by splining between the two open squares located at
$\tilde{\rho}_0^H$ and $\tilde{\rho}_0^Q$.
The resulting coexistence points (filled circles) 
are located at $\rho_1$ and $\rho_2$;
they are connected by the associated common tangent (the Maxwell line).
The spinodal boundary densities $\rho_A$ and $\rho_B$
are indicated by the open circles.
}\label{f:f-c}
\end{figure}

Figure \ref{f:f-c} shows $f_{T=0}^H(\rho)$ and $f_{T=0}^Q(\rho)$.
The former curve starts at zero but grows more rapidly
than the latter which starts at $B$, so the two curves cross and,
furthermore, since they both have positive curvature, a common tangent exists.
Because of this generic feature, it is expected that cold matter exhibits
a first-order phase transition when compressed.

While the simple gas models employed presumably
provide reasonable (though still somewhat idealized) descriptions of
the two individual phases well away from the coexistence region,
neither one is suitable in the phase-coexistence region.
In order to describe the transition region,
we represent the free energy density there 
by a $5^{\rm th}$-order polynomial $\tilde{f}_T(\rho)$
that matches the values of $f_T^H$, $\del_\rho f_T^H$, and $\del_\rho^2 f_T^H$
at a density $\tilde{\rho}_T^H\lesssim\rho_T^H$
and the values of $f_T^Q$, $\del_\rho f_T^Q$, and $\del_\rho^2 f_T^Q$
at a density $\tilde{\rho}_T^Q\gtrsim\rho_T^Q$,
where $\rho_T^H$ and $\rho_T^Q$ are those densities at which 
the two idealized curves $f_T^H(\rho)$ and $f_T^Q(\rho)$ have a common tangent.
For lower densities, $\rho\leq\tilde{\rho}_T^H$, 
we use the idealized hadron gas, $f_T^H(\rho)$,
and at higher energies, $\rho\geq\tilde{\rho}_T^Q$, 
we use the idealized plasma, $f_T^Q(\rho)$.

When the lower matching density $\tilde{\rho}_T^H$ 
is chosen sufficiently close to $\rho_T^H$
and the higher matching density $\tilde{\rho}_T^Q$ 
is chosen sufficiently close to $\rho_T^Q$
then the resulting spline function $\tilde{f}_T(\rho)$ also has a common
tangent and so the system has a first-order transition 
at the particular temperature $T$.
The associated densities (where the common tangent touches $\tilde{f}_T(\rho)$)
are then the coexistence densities, $\rho_1$ and $\rho_2$, at that temperature.

\begin{figure}          %       -----------------------------------------
\includegraphics[angle=0,width=3.1in]{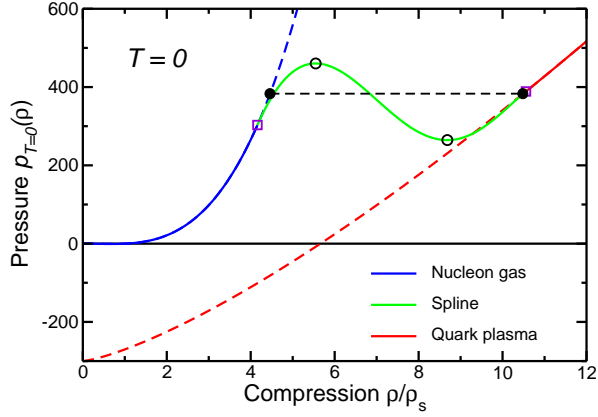}	%	phase crossing
\caption{The pressure at zero temperature, $p_{T=0}(\rho)$,
as a function of the degree of compression $\rho/\rho_c$,
corresponding to the free-energy density shown in Fig.\ \ref{f:f-c}.
The two individual phases are shown together with the connection between them
obtained by splining between the two open squares.
The resulting coexistence points (open circles)
are connected by the associated Maxwell line.
The spinodal boundaries are situated at the two extrema (open circles).
}\label{f:p-c}
\end{figure}

This spline procedure is illustrated in Fig.\ \ref{f:f-c} for $T=0$
and the corresponding pressure is shown in Fig.\ \ref{f:p-c}.
By suitable adjustment of the matching densities 
$\tilde{\rho}_T^H$ and $\tilde{\rho}_T^Q$,
it is thus possible to design an equation of state 
having the desired phase structure, namely a first-order phase transition
that becomes ever weaker as the temperature is raised
and terminates in a critical point at a finite density $\rho_c$.
The resulting phase coexistence boundaries, $(\rho_1,T)$ and $(\rho_2,T)$,
are shown in Fig.\ \ref{f:rho-Tx},
the full phase diagram having already been shown in Fig.\ \ref{f:T-c}.

            %       -----------------------------------------

            %       -----------------------------------------

%==============================================================================
\section{Spline method}
\label{spline}

We describe here a convenient spline method that enables us to match values 
and derivatives at two points. 

We seek an polynomial expression for the function $f(x)$
(here the free energy density $f_T(\rho)$)
that matches the specified values $a_0\equiv f(x=a)$ and $b_0\equiv f(x=b)$
as well as the associated derivatives up to any order $n$,
$a_i\equiv f^{(i)}_a\equiv (d^if/dx^i)_{x=a}$ and 
$b_i\equiv f^{(i)}_b\equiv (d^if/dx^i)_{x=b}$, $i=1,\dots,n$.

If $f_{n-1}(x)$ denotes the spline approximation that matches derivatives 
up to the $(n-1)^{\rm th}$ order (which is a polynomial of order $2n-1$),
then we may obtain the approximation for the next order $n$ by writing
\beq
f_n(x) = f_{n-1}(x)+(b-x)^n(x-a)^n
	{(b-x)\alpha_n+(x-a)\beta_n \over b-a}.
\eeq
Since the factor $(b-x)^n(x-a)^n$ and its derivatives up to order $n-1$
vanish at $a$ and $b$, it follows that
$f_n(x)$ satisfies the matching conditions up to order $n-1$, \ie\
$f_n^{(i)}(a)=f_{n-1}^{(i)}(a)=a_i$ and $f_n^{(i)}(b)=f_{n-1}^{(i)}(b)=b_i$
for $i<n$.

The only remaining task is thus to determine the two coefficients 
$\alpha_n$ and $\beta_n$ which can be done 
by matching also the $n^{\rm th}$ derivative at $a$ and $b$,
namely $f^{(n)}_a\doteq a_n$ and $f^{(n)}_b\doteq b_n$.
It is elementary to show that they are given by the following expressions,
\beq
\alpha_n={a_n-f_{n-1}^{(n)}(a) \over n!(b-a)^n}\ ,\,\
 \beta_n={b_n-f_{n-1}^{(n)}(b) \over n!(a-b)^n}\ .
\eeq

It is possible to also determine the derivatives of the spline function
by iteration.
In particular, the derivatives at the two matching points are given by
\beqar
&~& f_\nu^{(n)}(a)\ =\ f_{\nu-1}^{(n)}(a)\ +\\ &~&  \nonumber
	{n!\,\nu!(-)^{n-\nu}\over(n-\nu)!\,(2\nu-n+1)!}	
	[(\nu+1)\alpha_\nu-(n-\nu)\beta_\nu]\ ,\\
&~& f_\nu^{(n)}(b)\ =\ f_{\nu-1}^{(n)}(b)\ +\\ &~&  \nonumber
	{n!\,\nu!(-)^\nu\over(n-\nu)!\,(2\nu-n+1)!}	
	[(\nu+1)\beta_\nu-(n-\nu)\alpha_\nu]\ ,
\eeqar
where the first terms are present only for $n<2\nu$,
while the second terms are present only for $\nu\leq n\leq2\nu+1$.
We note that
\beq
f_n^{(2n+1)}(a) = (2n+1)!(-)^n[\beta_n-\alpha_n] = f_n^{(2n+1)}(b)\ ,
\eeq
consistent with the fact that the highest non-zero derivative is a constant.

The above expressions can be used iteratively to determine
the spline polynomial $f_n(x)$ for any oder $n\geq0$
as well as all of its $2n+1$ derivatives.

%%%%%%%%%%%%%%%%%%%%%%%%%%%%%%%%%%%%%%%%%%%%%%%%%%%%%%%%%%%%%%%%%%%%%%%%%%%%%%%
%                       REFECENCES:
%%%%%%%%%%%%%%%%%%%%%%%%%%%%%%%%%%%%%%%%%%%%%%%%%%%%%%%%%%%%%%%%%%%%%%%%%%%%%%%

%==============================================================================
			\end{document}